\documentclass[entropy,article,submit,moreauthors,pdftex]{mdpi}   
\usepackage{amsmath}
\firstpage{1} 
\pubvolume{1}
\issuenum{1}
\articlenumber{0}
\pubyear{2021}
\copyrightyear{2020}
\datereceived{} 
\dateaccepted{} 
\datepublished{} 
\hreflink{https://doi.org/} % If needed use \linebreak
\nolinenumbers
\Title{Effect of ergodic and non-ergodic fluctuations  on a charge diffusing in a stochastic Magnetic Field}

\TitleCitation{Particle in stochastic magnetic field}

\Author{Gerardo Aquino $^{1,\ddagger}$
Kristopher J. Chand\'ia $^{2,\ddagger}$ and Mauro Bologna $^{2,\ddagger}$*}

\AuthorNames{Firstname Lastname, Firstname Lastname and Firstname Lastname}

\AuthorCitation{Aquino, G.; Chand\'ia, K. J.; Bologna, M.}
\address{%
$^{1}$ \quad University of London; aquigerardo@gmail.com\\
$^{2}$ \quad Departamento de Ingenier\'ia El\'ectrica-Electr\'onica, Universidad de Tarapac\'a, Arica, Chile; kchandia@gmail.com;mauroh69@gmail.com}

\corres{Correspondence: mauroh69@gmail.com}

\secondnote{These authors contributed equally to this work.}
\abstract{In this paper, we study the basic problem of a charged particle in a stochastic magnetic field. We consider dichotomous fluctuations of the magnetic field  {where the sojourn time in one of the two states are distributed according to a given waiting time distribution either with Poisson or non-Poisson statistics, including as well the case of distributions with diverging mean time between changes of the field}, corresponding to an ergodicity breaking condition. We provide analytical and numerical results for all cases evaluating the average and the second moment of the position and velocity of the particle. We show that the field fluctuations induce diffusion of the charge with either normal or anomalous properties, depending on the statistics of the fluctuations, with distinct regimes from those observed, e.g., in standard Continuous Time Random Walk models.}


\begin{document}
%%%%%%%%%%%%%%%%%%%%%%%%%%%%%%%%%%%%%%%%%%
\setcounter{section}{0} %% Remove this when starting to work on the template.
\nolinenumbers
\section{introduction}\label{introduction}
Diffusive processes occur in many physical, chemical and engineering applications. When the diffusion process is taking place, the quantities related to the spreading species take random values. Since Einstein and Smoluchowski's work on Brownian motion~\cite{albert,smo}, diffusive phenomena have been a fundamental subject of intense research. Both derivations (Einstein and Smoluchowski's) lead to the well-known diffusion relationship, in the one-dimensional case, $\langle x^2\rangle= 2 Dt$, with $D$ the diffusion coefficient. 
Several relevant physical and biological phenomena have been discovered in the last few decades, showing an anomalous relationship between mean squared displacement and time, $\langle x^2\rangle\propto t^\alpha$. For example, diffusion through porous media or within a crowded cellular environment, making anomalous diffusion a relevant subject of research work~\cite{bel,bou,aq,zas} and~\cite{met,mainardi} for a review.

This paper presents a detailed study of a particle moving in a fluctuating magnetic field that is directly connected to plasma physics. In particular, it is relevant in many technological applications such as, for example, plasma confinement~\cite{shun}. We focus on the diffusion of the particles caused by the magnetic field fluctuations, which can destroy the plasma confinement. The fluctuations of physical quantities are an almost inevitable occurrence and generate diffusion processes of the related quantity~\cite{mitt,marc,a}. It is important, therefore, to have an adequate theoretical framework to model their effect. We aim to fill the gap relative to the case of diffusion in a fluctuating magnetic field, which, to our knowledge, has not yet been explored and modeled so far in the case of non-ordinary statistics. 

The paper is organized as follows: In Sec.~\ref{magnetic_force} we introduce and formally define the problem for generic fluctuations.
In Sec.~\ref{dichotomous_processes}, we introduce the case of dichotomous fluctuations, and we afford a complete analytical and numerical treatment of Poisson and non-Poisson statistics.
In Sec.~\ref{sec_diffusion}, we look at the average squared displacement and, starting from the Poisson case, we move to non-Poisson, power-law distributed fluctuations
exploring, therefore also the non-ergodic regime when the average time of the fluctuations diverges ( {see Ref.~\cite{eli} for an extended discussion, and Ref.~\cite{ralf} and references therein for a review and a historical perspective}). The diffusion properties in this regime are characterized
by the analytical and numerical derivation of the average squared displacement. Secs.~\ref{sec_conclusions} and~\ref{sec_conclusions2} summarize results and draw final conclusions.

%The introduction  defines the purpose of the work and its significance. The current state of the research field should be reviewed carefully and key publications cited. Please highlight controversial and diverging hypotheses when necessary. Finally, briefly mention the main aim of the work and highlight the principal conclusions. As far as possible, please keep the introduction comprehensible to scientists outside your particular field of research. Citing a journal paper \cite{ref-journal}. Now citing a book reference \cite{ref-book1,ref-book2} or other reference types \cite{ref-unpublish,ref-communication,ref-proceeding}. Please use the command \citep{ref-thesis,ref-url} for the following MDPI journals, which use author--date citation: Administrative Sciences, Arts, Econometrics, Economies, Genealogy, Histories, Humanities, IJFS, Journal of Intelligence, Journalism and Media, JRFM, Languages, Laws, Religions, Risks, Social Sciences.
 
%%%%%%%%%%%%%%%%%%%%%%%%%%%%%%%%%%%%%%%%%%

%The remainder of this paper The paper is organized as follows: In Sec.~\ref{magnetic_force}

\section{Stochastic equation for the magnetic force}\label{magnetic_force}
Let us consider the following classical equation~\cite{land8,jak}

\begin{eqnarray}  \label{f0}
m \frac{d \mathbf{v}}{dt}= q \left(\mathbf{v} \times \mathbf{B}+ \mathbf{E}\right)
\end{eqnarray}
where~$m$ is the mass of the charge $q$, ~$ \mathbf{v}$ is the velocity,~$ \mathbf{B}$ the magnetic field, and~$ \mathbf{E}$ the electric field. We focus on the case where the particle is traveling in a region with a uniform magnetic field randomly fluctuating, i.e.,~$ \mathbf{B}= \mathbf{B}_0+ \mathbf{B}_1(t)$ where~$\mathbf{B}_1(t)=B_1\text{\boldmath$\xi$}(t)$ with~$\text{\boldmath$\xi$}(t)$ the stochastic fluctuation of the magnetic field, where $\text{\boldmath$\xi$}(t)=\xi(t)\mathbf{n}_{\xi}$.  {We shall consider a dichotomous fluctuation, with values of $\pm 1$ where the sojourn time in one of the two states are distributed according to the distribution~$\psi(t)$. We will consider the case when~$\psi(t)$ is an exponential, $\psi(t)=\gamma\exp[-\gamma t]$, or Poisson case, and the case when $\psi(t)$ is a power law, $\psi(t)\propto t^{-\alpha-1}$ with $0<\alpha<2$, or non-Poisson case.}

The main reason for choosing dichotomous fluctuations rests on the fact that taking the finite values fluctuations may better represent a physical system. Also, we stress that in  the appropriate limit,  the most well-known noises in literature, such as the gaussian white noise and the white shot noise are recovered within this framework ~\cite{van,cal}. Further, we consider the case where no external electric field is applied. Taking the magnetic field in the~$z$ direction, $\mathbf{B}\equiv B(t)\mathbf{k}=[B_0+ B_1\xi(t)]\mathbf{k}$, and using cartesian coordinates, we have

\begin{eqnarray} \label{f1}
&&\frac{d v_x}{dt}=   \omega(t) v_y+ \frac{q}{m} E_x 
\\\label{f2}
&&\frac{d v_y}{dt}=-\omega(t) v_x+  \frac{q}{m} E_y
\end{eqnarray}
with $ \omega(t)=qB(t)/m$ the time-dependent Larmor frequency and $E_x$,~$E_y$ the induced electrical field. As a further simplification, we consider the time scale of the fluctuation much larger than the time associated with the unperturbed Larmor frequency $\omega_0=qB_0/m$.  {It is worthy to note that Eqs.~(\ref{f1}), (\ref{f2}) can be reduced to a second order stochastic differential equation for the function $z=x+ i y$. The resulting equation has a strong similarity with the mechanical system studied in Refs.~\cite{gitt2,gitt} where  the authors study a linear damped oscillator with a noise perturbing both the oscillator mass and the friction. A detailed study of this equation is out of the scope of this paper and it is left to an upcoming publication.}

Neglecting the induced electrical field, $ \mathbf{E}\approx 0$, (see also~\cite{bol}), we have 

\begin{eqnarray} \label{f1e0}
&&\frac{d v_x}{dt}=   \omega(t) v_y
\\\label{f2e0}
&&\frac{d v_y}{dt}=-\omega(t) v_x
\end{eqnarray}
For sake of completeness, we end this section showing the equation for the density probability $P(v_x.v_y,t)=P(\mathbf{v},t)$ associated to the stochastic equations~(\ref{f1e0}), (\ref{f2e0}). To obtain a closed equation for $P(\mathbf{v},t)$ we will assume that $\xi$ is a Poisson process although, in the next sections, the analysis of the relevant quantiles, $\langle v_x(t)\rangle$, $\langle v_y(t)\rangle$, $\langle x(t)^2+y(t)^2\rangle$, will include also non-Poisson processes. Using the Liouville approach we write the following continuity equation 

\begin{eqnarray} \label{liouv}
\frac{\partial \rho}{\partial t}= -\nabla\cdot\left[\rho\mathbf{v}\times\boldsymbol\omega\right]=
 -\nabla\cdot\left[\rho\mathbf{v}\times\boldsymbol\omega_0\right] -\nabla\cdot\left[\xi\rho\mathbf{v}\times\boldsymbol\omega_1\right]
\end{eqnarray}
where for sake of compactness we dropped the function arguments and we introduced the symbols $$\nabla\equiv \frac{\partial}{\partial v_x}\mathbf{i}+\frac{\partial  }{\partial v_y}\mathbf{j},\,\,\boldsymbol\omega= \left[\frac{q B_0}{m}+  \frac{q B_1}{m}\xi(t)\right]\mathbf{k}\equiv\boldsymbol\omega_0+\boldsymbol\omega_1\xi(t).
$$
The stochastic density~$\rho$ is related to $P(v_x.v_y,t)$ via the Van Kampen's lemma~\cite{van2} $\langle \rho\rangle = P(\mathbf{v},t)$ where the average is performed on the $\xi$ realizations. Also, we will use the Shapiro -Loginov formulae of differentiation~\cite{log}

\begin{equation}\label{shap}
\frac{\partial}{\partial t}\langle\xi(t)\rho(t)\rangle=
-\gamma\langle\xi(t)\rho(t)\rangle+\langle\xi(t)\frac{\partial}{\partial
t}\rho(t)\rangle,
\end{equation}
that holds true for processes with a $n$th correlation
function fulfilling the condition

$$\frac{\partial}{\partial
t}\langle\xi(t)\xi(t_1)\cdots\xi(t_n)\rangle
=-\gamma\langle\xi(t)\xi(t_1)\cdots\xi(t_n)\rangle.$$ 
 {In particular, this applies to Poisson, Gaussian and Markov jump processes, with a correlation function given by $\langle\xi(t_1)\xi(t_2)\rangle\sim \exp[-\gamma |t_2-t_1|]$}. Taking the average of Eq.~(\ref{liouv}), defining $P_1(\mathbf{v},t)\equiv \langle \xi\rho\rangle$ and using Eq.~(\ref{shap}), we may write the system

\begin{eqnarray} \label{liouv2}
&&\frac{\partial P(\mathbf{v},t)}{\partial t}=
 -\mathbf{v}\times\boldsymbol\omega_0\cdot\nabla P -\mathbf{v}\times\boldsymbol\omega_1\cdot\nabla P_1,
 \\\label{liouv3}
 &&\frac{\partial P_1(\mathbf{v},t) }{\partial t}=-\gamma P_1 \times\boldsymbol\omega_0\cdot\nabla P_1+\mathbf{v}\times\boldsymbol\omega_1\cdot\nabla P
\end{eqnarray}
Taking the time derivative of Eq.~(\ref{liouv2}), and after some algebra we obtain the following equation for the probability density

\begin{eqnarray}\nonumber
\frac{\partial^2 P(\mathbf{v},t)}{\partial t^2}&=&-\gamma \frac{\partial P}{\partial t}-\gamma \mathbf{v}\times\boldsymbol\omega_0\cdot\nabla P+ \mathbf{v}\times\boldsymbol\omega_0\cdot\nabla \left[\mathbf{v}\times\boldsymbol\omega_0\cdot\nabla P\right]+
\\\label{liouv4}
&& -
 \mathbf{v}\times\boldsymbol\omega_1\cdot\nabla \left[\mathbf{v}\times\boldsymbol\omega_1\cdot\nabla P\right].
\end{eqnarray}

\section{Dichotomous processes}\label{dichotomous_processes}
As stated in Sec.~\ref{magnetic_force}, in this section we will consider a magnetic field with a fluctuating component which is assumed to be dichotomous. 
 Dichotomous fluctuations have the nice property of  assuming finite values but, despite their relative simplicity,  they can be shown to allow one to recover both gaussian white noise and white shot noise~\cite{van} within an appropriate limit  procedure. Formally the exact solution of Eqs.~(\ref{f1e0}), (\ref{f2e0}) is

\begin{eqnarray}   \label{f3}
&&v_x(t)=v_0\sin \left[\int_0^t\!\!\omega(u) du\right] = v_0\mathrm{Im}\left[\exp \left[i \omega_0 t\right]
\exp \left[i \omega_1\int_0^t \!\!\xi(u)du\right]\right]\!,
\\ \label{f4}
&&v_y(t)=v_0\cos \left[\int_0^t\!\!\omega(u) du\right] = v_0\mathrm{Re}\left[\exp \left[i \omega_0 t\right]
\exp \left[i \omega_1\int_0^t \!\!\xi(u)du\right]\right]\!,
\end{eqnarray}
where~$v_0$ is the initial velocity along the $y$ axis. For the average we have

\begin{eqnarray}   \label{f5}
&&\langle v_x(t)\rangle= v_0\mathrm{Im}\left[\exp \left[i \omega_0 t\right]
\Big\langle\exp \left[i \omega_1\int_0^t \!\!\xi(u)du\right]\Big\rangle\right],
\\\label{f6}
&&\langle v_y(t)\rangle= v_0\mathrm{Re}\left[\exp \left[i \omega_0 t\right]
\Big\langle\exp \left[i \omega_1\int_0^t \!\!\xi(u)du\right]\Big\rangle\right].
\end{eqnarray}
As we infer from Eqs.~(\ref{f5}), (\ref{f6}), we need to evaluate the average of the exponential of the noise integral. 
For this purpose we consider the stochastic equation

\begin{equation}\label{campo}
\frac{dU}{dt} =\xi(t)
\end{equation}
with $\xi(t)$ a dichotomous fluctuation where the sojourn time in one of the two states
are distributed according to the distribution~$\psi(t)$. We assume that the event, occurring at each random time~$t_i$, changes the $\xi(t)$ sign. 
 {When these events  occur with a constant rate $\gamma$ this corresponds to a Poisson process, and is characterized by an exponential distribution.
As stated in the Introduction, we will consider here both the case of exponential (Poisson) distribution  with 
\begin{equation}
\psi(t)=\gamma \exp[-\gamma t]
\end{equation}
 and the case of non-Poisson process with power-law distribution, characterized  by the following asymptotic behavior
 \begin{equation}
\psi(t)\propto\left( \frac{t }{ T}\right)^{-\alpha-1},\,\,\,\,{t\gg T}
\end{equation}
with $0<\alpha<2$, which corresponds to a regime with a diverging second  moment ($1<\alpha<2$)  and  diverging first and second moment ($0< \alpha<1$).  The latter case, characterized by the absence of a finite time scale, corresponds to a condition of ergodicity breaking.\newline}
The formal solution of Eq.~(\ref{campo}) is

\begin{equation}
U(t)=\int_0^t\xi(u)du.
\end{equation}
Considering the formal solution for the velocity of the charge, Eqs.~(\ref{f5}) and (\ref{f6}), we need to evaluate the 
quantity~$\Big\langle\exp \left[i \omega_1\int_0^t\xi(u)du\right]\Big\rangle$. For this purpose we use the exact formula in the Laplace representation~\cite{aq2,aq2bis} (see Ref.~\cite{ref-thesis} for detailed calculations)

\begin{eqnarray}\nonumber
&&\mathcal{L}\left[\Big\langle\exp \left[i \omega_1\int_0^t \!\!\xi(u)du\right]\Big\rangle\right]=
\frac{1}{2} \left[\frac{\left(1+\hat{\psi} \left(s-i \omega _1\right)\right) \hat{\Psi} \left(s+i \omega _1\right)}{1-\hat{\psi} \left(s-i \omega _1\right) \hat{\psi}\left(s+i \omega _1\right)}+\right.
\\\label{general}
&&\left.\frac{\left(1+\hat{\psi} \left(s+i \omega _1\right)\right) \hat{\Psi} \left(s-i \omega _1\right)}{1-\hat{\psi}\left(s-i \omega _1\right)\hat{\psi} \left(s+i \omega _1\right)}\right]=
\mathrm{Re} \left[\frac{\left(1+\hat{\psi} \left(s+i \omega _1\right)\right)\hat{\Psi} \left(s-i \omega _1\right)}{1-\hat{\psi} \left(s-i \omega _1\right) \hat{\psi}\left(s+i \omega _1\right)}\right]
\end{eqnarray}
where $\hat{\psi}(s)$ is the Laplace transform of $\psi(t)$ and $ \Psi \left(t\right)$ (and consequently $\hat{\Psi}(s)$ is its Laplace transform) is the probability that no switch occurs for a generic interval of time $t$, i.e.

\begin{equation}
\Psi \left(t\right)=1-\int_0^t\psi \left(t_1\right)dt_1=\int_t^\infty\psi \left(t_1\right)dt_1
\end{equation}
The Poisson case does not present difficulties and, in the time representation,  gives the expression

\begin{eqnarray}\nonumber
&&\Big\langle\exp \left[i \omega_1\int_0^t \!\!\xi(u)du\right]\Big\rangle=
\\\label{average_eq}
&&\exp\left[-\frac{\gamma}{2}   t\right] \left[\frac{\gamma  \sinh \left(t \sqrt{\frac{\gamma ^2}{4}-\omega_1 ^2}\right)}{2\sqrt{\frac{\gamma ^2}{4}-\omega_1 ^2}}+ \cosh \left(t \sqrt{\frac{\gamma ^2}{4}-\omega_1^2}\right)\right].
\end{eqnarray}
We are now in position to write a closed expression for Eqs.~(\ref{f5}), (\ref{f6}). The average of the velocity components is

\begin{eqnarray}   \nonumber
&&\langle v_x(t)\rangle= v_0\sin \left[\omega_0 t\right]
\\ \label{f5b}
&&\exp\left[-\frac{\gamma}{2}   t\right] \left[\frac{\gamma  \sinh \left(t \sqrt{\frac{\gamma ^2}{4}-\omega_1 ^2}\right)}{2\sqrt{\frac{\gamma ^2}{4}-\omega_1 ^2}}+ \cosh \left(t \sqrt{\frac{\gamma ^2}{4}-\omega_1^2}\right)\right]
\\\nonumber
&&\langle v_y(t)\rangle= v_0\cos \left[\omega_0 t\right]
\\ \label{f5bis}
&&\exp\left[-\frac{\gamma}{2}   t\right] \left[\frac{\gamma  \sinh \left(t \sqrt{\frac{\gamma ^2}{4}-\omega_1 ^2}\right)}{2\sqrt{\frac{\gamma ^2}{4}-\omega_1 ^2}}+ \cosh \left(t \sqrt{\frac{\gamma ^2}{4}-\omega_1^2}\right)\right]
\end{eqnarray}
We now study Eq.~(\ref{general}) for the non-Poisson  {case with a power-law distribution of the type of  Eq. (16) with $0<\alpha<2$} where the non-ergodicity of the process plays an important role. Some difficulties arise in inverting the Laplace transform, in particular in the region defined by $1<\alpha<2$. This is partially due to the fact that, while for $0<\alpha<1$ the calculation in the Laplace transform can be carried out using as waiting-time distribution the derivative of the Mittag-Leffler function (see for example Ref.~\cite{mainardi} and references therein), for the region $0<\alpha<2$ the Laplace transform is usually a complicated function, and the inversion of the final result is not an easy task. Traditionally the inversion of a Laplace transform for a large value of time t is performed using the Tauberian theorem, i.e., taking the development for the Laplace parameter $s\to 0$. If the function to invert is a hard-to-handle function, it is not always clear where to stop the development (see \cite{bol3} for detailed examples). To overcome this difficulty, we may use as waiting-time distribution \cite{bol4}

\begin{equation}\label{nd8}
\psi(t) =\frac{\sin \left(\frac{\pi  \alpha}{2}\right) \cos  \left(\frac{t}{T}\right) +\cos \left(\frac{\pi \alpha}{2}\right) \sin \left(\frac{t}{T}\right)}{T\cos \left(\frac{\pi \alpha}{2}\right)}-\frac{\sin \left(\frac{\pi \alpha}{2}\right) \cos_\alpha \left(\frac{t}{T}\right)+\cos \left(\frac{\pi\alpha}{2}\right) \sin_\alpha  \left(\frac{t}{T}\right)}{T\cos \left(\frac{\pi \alpha}{2}\right)}.
\end{equation}
where $T$ is a time-scale parameter and, by definition~\cite{mau,west},

\begin{eqnarray}\label{nd6}
\cos_\alpha t\equiv  \frac{E_{\alpha}^{it}+E_{\alpha}^{-it}}{2},\,\,\sin_\alpha t\equiv  \frac{E_{\alpha}^{it}-E_{\alpha}^{-it}}{2i}
\end{eqnarray}
and

\begin{equation}\label{nd4}
E_{\alpha}^{t}\equiv D_{t}^{\alpha}\exp[t]=\sum_{n=0}^{\infty}\frac{t^{n-\alpha}}{\Gamma(n+1-\alpha)},
\end{equation}
where $D_{t}^{\alpha}$ is the Riemann-Liouville fractional derivative.  {The functions $\cos_\alpha t$ and $\sin_\alpha t$ compensate the oscillatory behavior of the ordinary trigonometric functions, and what remains is a positive power law, i.e., $t^{-\alpha-1}$. A rigorous proof is given in Ref.~\cite{bol4}.} Despite its complicated structure in time representation, its Laplace transform is

\begin{equation}\label{nd11}
\hat{\psi}(s)=\frac{1+sT \tan \left(\frac{\pi  \alpha }{2}\right)-\sec \left(\frac{\pi  \alpha }{2}\right) (sT)^{\alpha }}{(sT)^2+1},\,\,
0<\alpha<2,\,\,\alpha\neq 1.
\end{equation}
For $\alpha=1$ we have to take the limit and we obtain

\begin{equation}\label{nd11lim}
\hat{\psi}(s)=\lim_{\alpha\to1}\frac{1+sT \tan \left(\frac{\pi  \alpha }{2}\right)-\sec \left(\frac{\pi  \alpha }{2}\right) (sT)^{\alpha }}{(sT)^2+1}=
\frac{1+ \frac{2 sT}{\pi}\log (sT)}{(sT)^2+1}.
\end{equation}
The proposed distribution has a simple structure based on power law and its validity is in the non-Poisson ranges $0<\alpha<2$. Using the property $\mathcal{L}\left[\exp[\pm i \omega_1 t] f(t)\right]=\hat{f}(s\mp i \omega_1)$, defining the new Laplace variables, $v=s- i \omega_1$ , we can reduce the inverse Laplace problem of Eq.~(\ref{general}) the following inversion Laplace transform

\begin{eqnarray}\label{general_bis}
\mathrm{Re} \left[\frac{\left(1+\hat{\psi} \left(v+2 i \omega _1\right)\right)\hat{\Psi} \left(v\right)}{1-\hat{\psi} \left(v+2 i \omega _1\right) \hat{\psi}\left(v\right)}\right].
\end{eqnarray}
Since we are interested in the asymptotic limit, taking the limit for $v\to 0$ 
we may invert Eq.~(\ref{general_bis}). We find for the asymptotic expression

\begin{eqnarray}\label{first}
\Big\langle\exp \left[i \omega_1\int_0^t \!\!\xi(u)du\right]\Big\rangle\approx 
A\frac{\exp\left[ i \omega_1 t +\phi\right] }{t^{\alpha}}
\end{eqnarray}
where $A$ and $\phi$ are constant depending on $\omega_1,T,\alpha$. In the region $0<\alpha<1$ there is no dependence 
on the parameter $T$ and we obtain

\begin{eqnarray}\nonumber
&&\Big\langle\exp \left[i \omega_1\int_0^t \!\!\xi(u)du\right]\Big\rangle\approx 
2 \mathrm{Re}\left[\exp (i \omega_1 t )E_{\alpha}\left[-(2 i \omega_1 t )^{\alpha }\right]\right]
\\\label{mu12_1}
&&\approx\frac{\sqrt{\pi } 
J_{\alpha -\frac{1}{2}}(  \omega_1 t )}{\Gamma (1-\alpha ) (2 \omega_1 t)^{\alpha -\frac{1}{2}}}
\end{eqnarray}
where $E_{\alpha}\left[-(2 i \omega_1 t )^{\alpha }\right]$ is the Mittag-Leffler function defined as

$$E_{\alpha}\left[z\right]=\sum_{n=0}^\infty\frac{z^n}{\Gamma(n\alpha+1)},$$
and $J_\nu(z)$ is the Bessel function of the first kind. From an asymptotic point of view, all the expressions contained in Eqs.~(\ref{first}) and (\ref{mu12_1}) have the same accuracy. 
The advantage of the expression written as in the last line of Eq.~(\ref{mu12_1}) is that in the case $\alpha=1/2$, it provides an exact expression, i.e.

\begin{eqnarray}   \nonumber
&&\langle v_x(t)\rangle= v_0\mathrm{Im}\left[\exp \left[i \omega_0 t\right]
\Big\langle\exp \left[i \omega_1\int_0^t \!\!\xi(u)du\right]\Big\rangle\right]=
\\ \label{np_f5}
&&v_0\sin\left[ \omega_0 t\right]
J_0 \left[ \omega_1 t\right],
\\\nonumber
&&\langle v_y(t)\rangle= v_0\mathrm{Re}\left[\exp \left[i \omega_0 t\right]
\Big\langle\exp \left[i \omega_1\int_0^t \!\!\xi(u)du\right]\Big\rangle\right]=
\\ \label{np_f6}
&&v_0\cos\left[ \omega_0 t\right]
J_0 \left[ \omega_1 t\right].
\end{eqnarray}
This can be directly checked using the distribution $P(x,t)$ derived by Lamperti~\cite{bol2,lamp}, and which describes the distribution 
associated to Eq.~(\ref{campo}) for $0<\alpha<1$. Integrating Eqs.~(\ref{np_f5}), (\ref{np_f6}) the result with respect to time, we obtain

\begin{eqnarray}\label{np_f7}
&&\langle x(t)\rangle=  v_0\int_0^t\sin\left[ \omega_0 t\right]
J_0 \left[ \omega_1 t\right]dt+x_0,
\\ \label{np_f8}
&&\langle y(t)\rangle= v_0\int_0^t\cos\left[ \omega_0 t\right]
J_0 \left[ \omega_1 t\right]dt+y_0
\end{eqnarray}

\begin{figure}[H]
\includegraphics[width=10 cm]{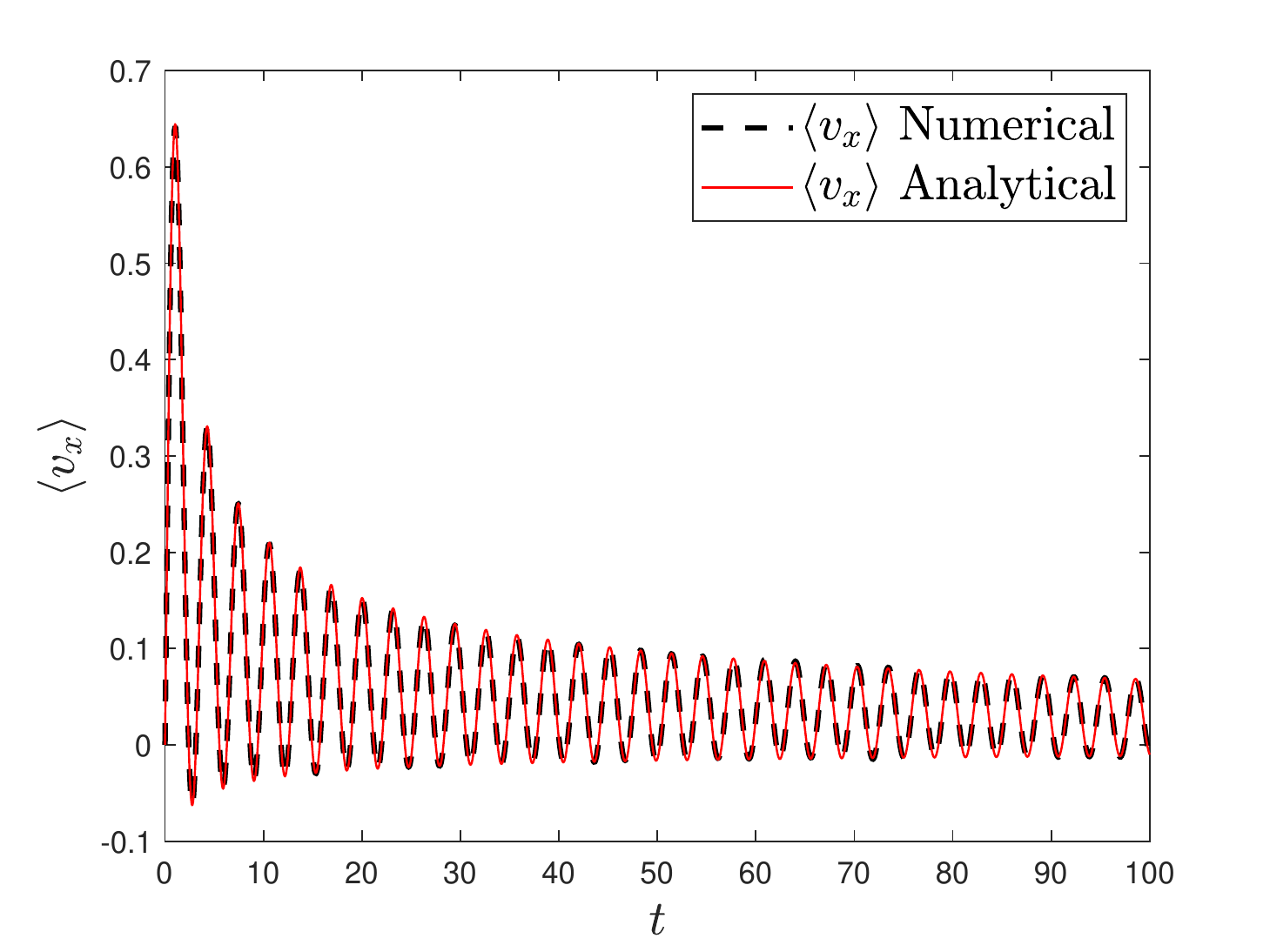}
\caption{Plot of the analytical (redline)  and the numerical solution of the average velocity along the $x$ axis with $\alpha=0.5$, $T=0.001$, $\omega_0=\omega_1=1$. 
Number of realizations $200$k.\label{fig1}}
\end{figure}

\begin{figure}[H]
\includegraphics[width=10 cm]{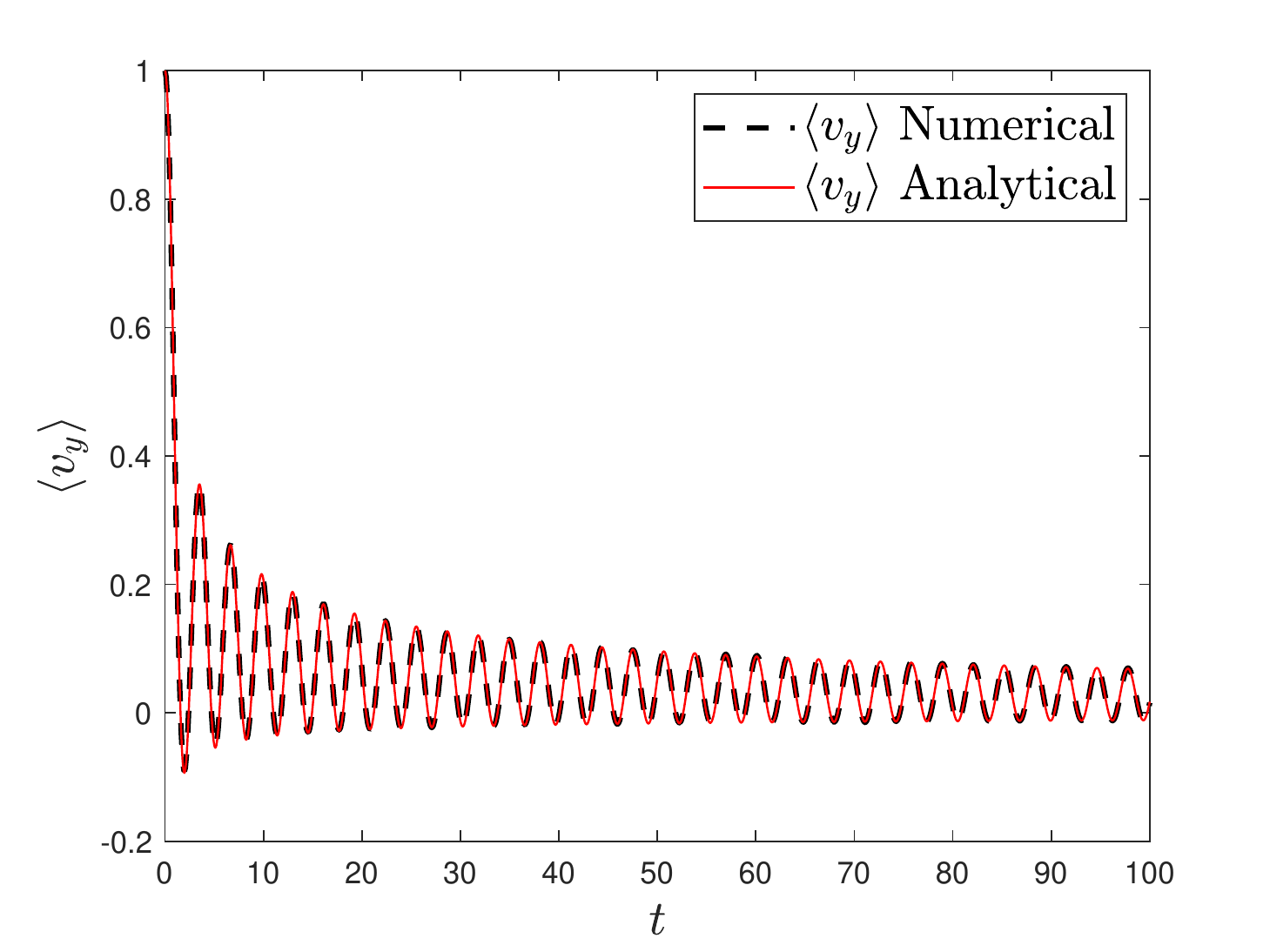}
\caption{Plot of the analytical (redline)  and the numerical solution of the average velocity along the $y$ axis with $\alpha=0.5$, $T=0.001$, $\omega_0=\omega_1=1$. Number of realizations $200$k. \label{fig2}}
\end{figure}

Asymptotically we have 
\begin{eqnarray}\label{np_f9}
&&x(\infty)=
\begin{cases}
\frac{v_0}{\sqrt{\omega_0^2-\omega_1^2}}+x_0, & \text{if }\omega_1 <\omega_0 ,\\
x_0, & \text{if }\omega_0 <\omega _1.
\end{cases}
\\\label{np_f10}
&&y(\infty)=\begin{cases}
\frac{v_0}{\sqrt{\omega _1^2-\omega_0 ^2}}, & \text{if }\omega_0 <\omega _1,\\
y_0, & \text{if }\omega_1<\omega_0.
\end{cases}
\end{eqnarray}
The resonant case $\omega_1=\omega_0$ generates diverging average positions

\begin{eqnarray}\label{np_f11}
&&\langle x(t)\rangle= 
v_0 t [\sin (\omega_0 t) J_0( \omega_0 t)- \cos ( \omega_0 t) J_1( \omega_0 t)]+x_0,
\\ \label{np_f12}
&&\langle y(t)\rangle=  v_0 t [\sin ( \omega_0 t) J_1(\omega_0 t)+\cos ( \omega_0 t) J_0( \omega_0 t)]+y_0.
\end{eqnarray}
 {Figures~\ref{fig1},~\ref{fig2} show the comparison between analytical results and numerical simulations for the quantities $\langle v_x(t)\rangle$ and $\langle v_y(t)\rangle$. Figures~\ref{figpoi1},~\ref{figpoi2},~\ref{fignopoi1},~\ref{fignopoi2} show single realizations of the stochastic trajectories, and Figures~\ref{fig3},~\ref{fig4} show the comparison between analytical results and numerical simulations for the quantities $\langle x(t)\rangle$ and $\langle y(t)\rangle$. Finally, Figure~\ref{figerror} shows the percent error as a function of the number of the realizations. The error decreases starting from $10\%$ (blu line, 10k realizations) to $< 1\%$ (yellow line, 200k realizations)}

 \begin{minipage}{\linewidth}
      \centering
      \begin{minipage}{0.45\linewidth}
          \begin{figure}[H]
              \includegraphics[width=\linewidth]{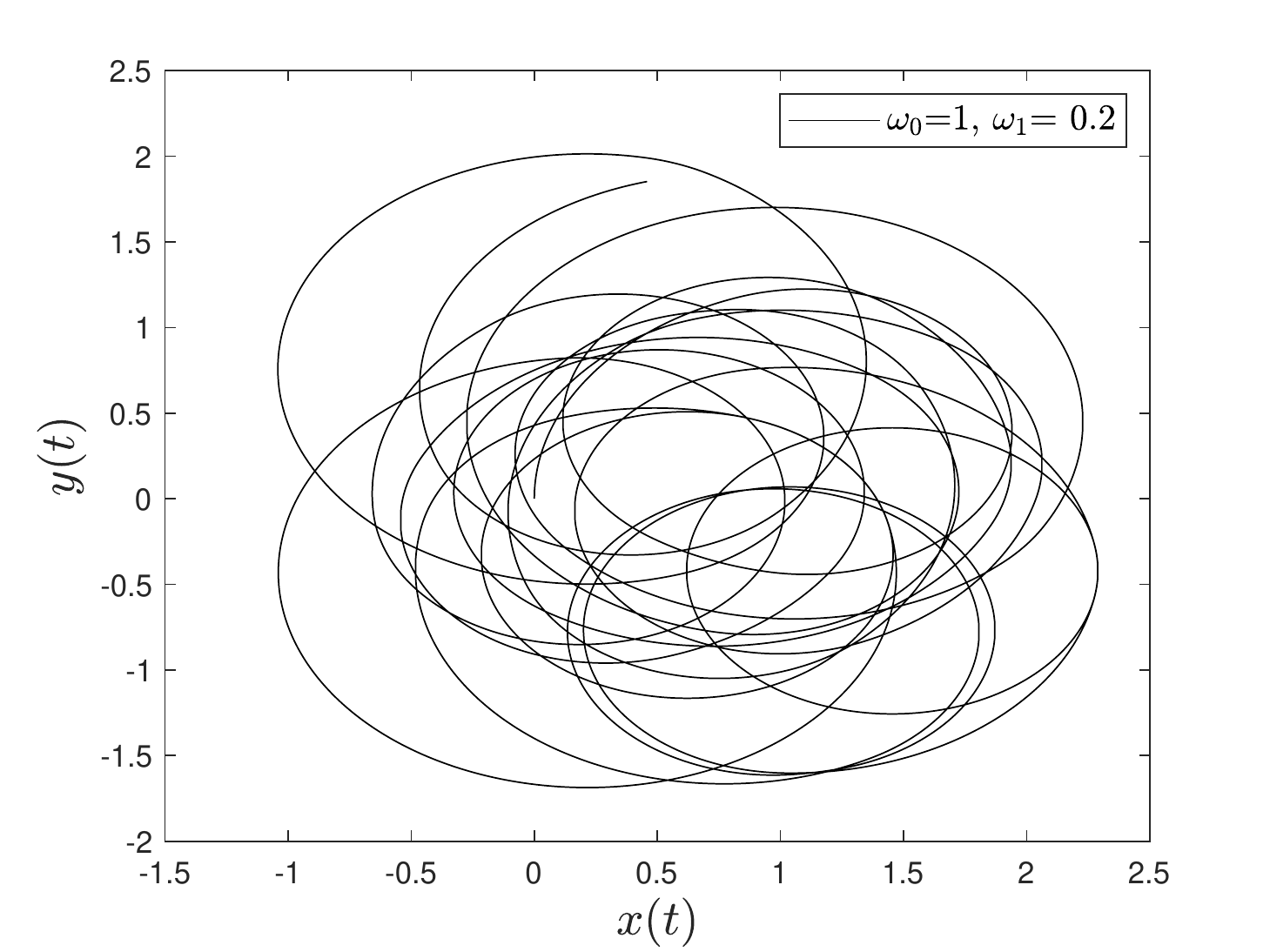}
              \caption{Single realization of the trajectory of a charge in the Poisson case. The values of the parameters are $\gamma=1$, 
              $\omega_0=1$, $\omega_1=0.2$ and $t=100$\label{figpoi1}}
          \end{figure}
      \end{minipage}
      \hspace{0.05\linewidth}
      \begin{minipage}{0.45\linewidth}
          \begin{figure}[H]
              \includegraphics[width=\linewidth]{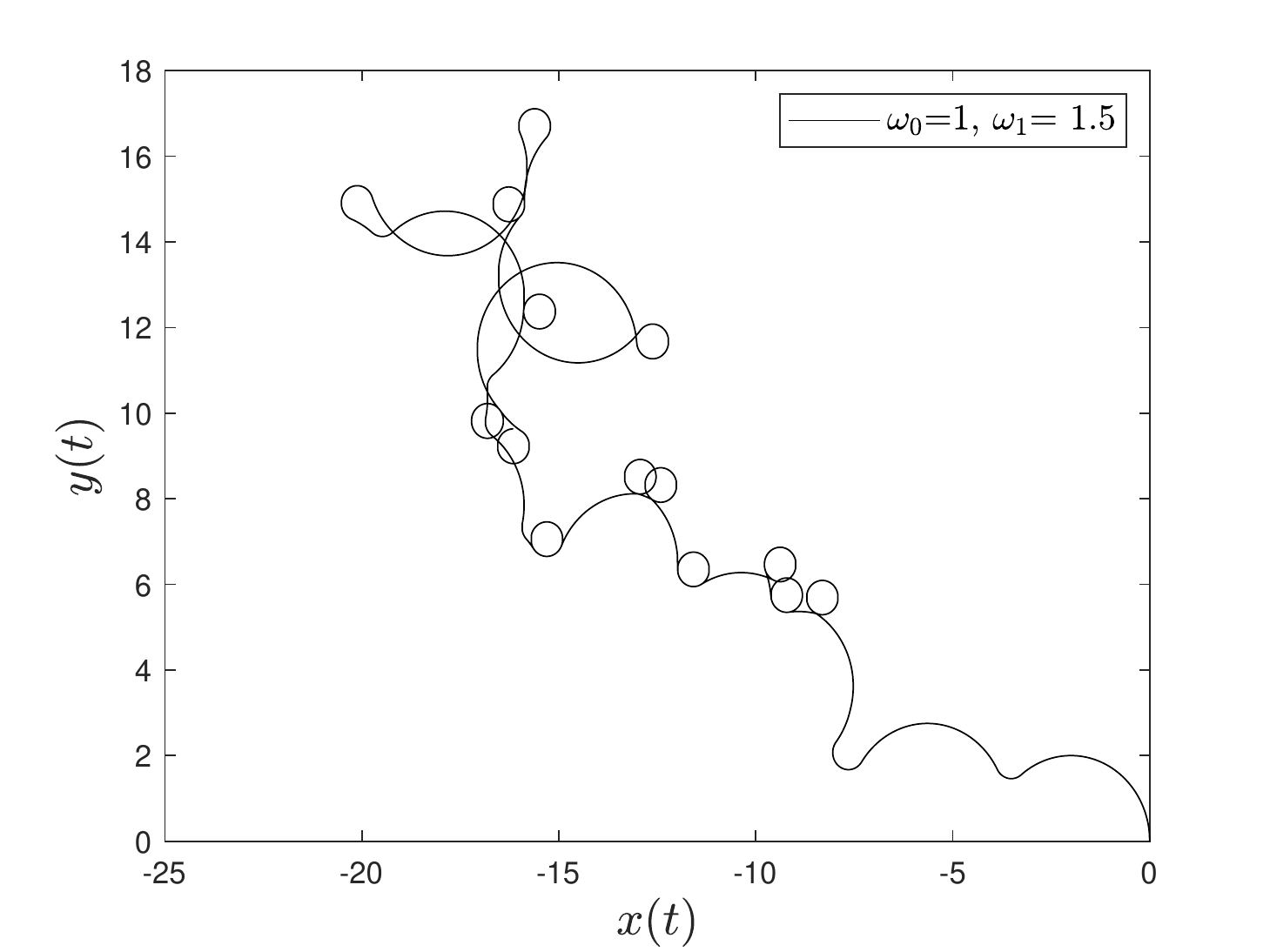}
              \caption{Single realization of the trajectory of a charge in the Poisson case. The values of the parameters are $\gamma=1$, 
              $\omega_0=1$, $\omega_1=1.5$ and $t=100$\label{figpoi2}}
          \end{figure}
      \end{minipage}
      \hspace{0.05\linewidth}
      \begin{minipage}{0.45\linewidth}
          \begin{figure}[H]
              \includegraphics[width=\linewidth]{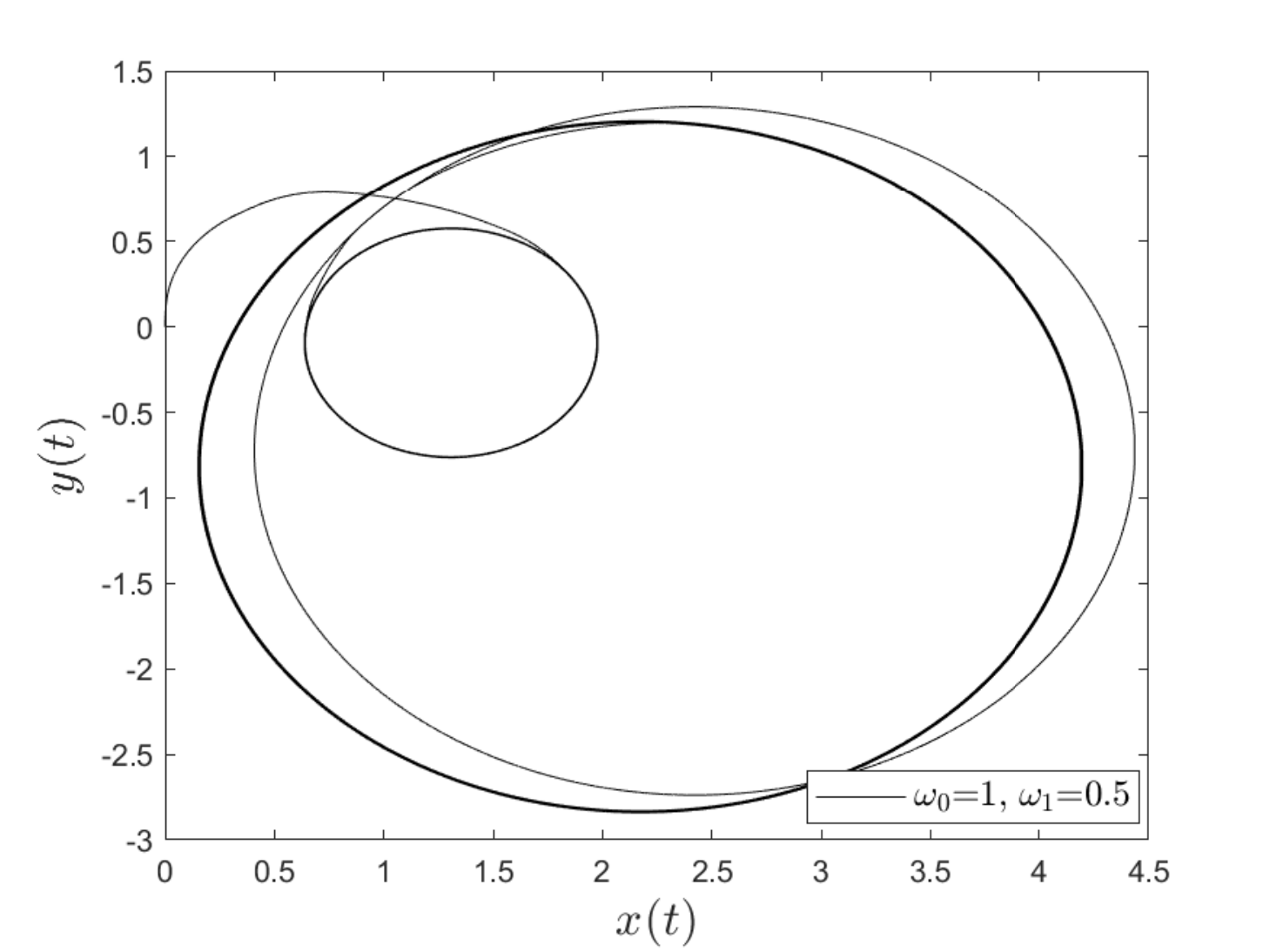}
              \caption{Single realization of the trajectory of a charge in the non-Poisson case. The values of the parameters are $\alpha=0.5$, $T=0.001$, $\omega_0=1$, $\omega_1=0.5$ and $t=500$\label{fignopoi1}}
          \end{figure}
      \end{minipage}
      \hspace{0.05\linewidth}
      \begin{minipage}{0.45\linewidth}
          \begin{figure}[H]
              \includegraphics[width=\linewidth]{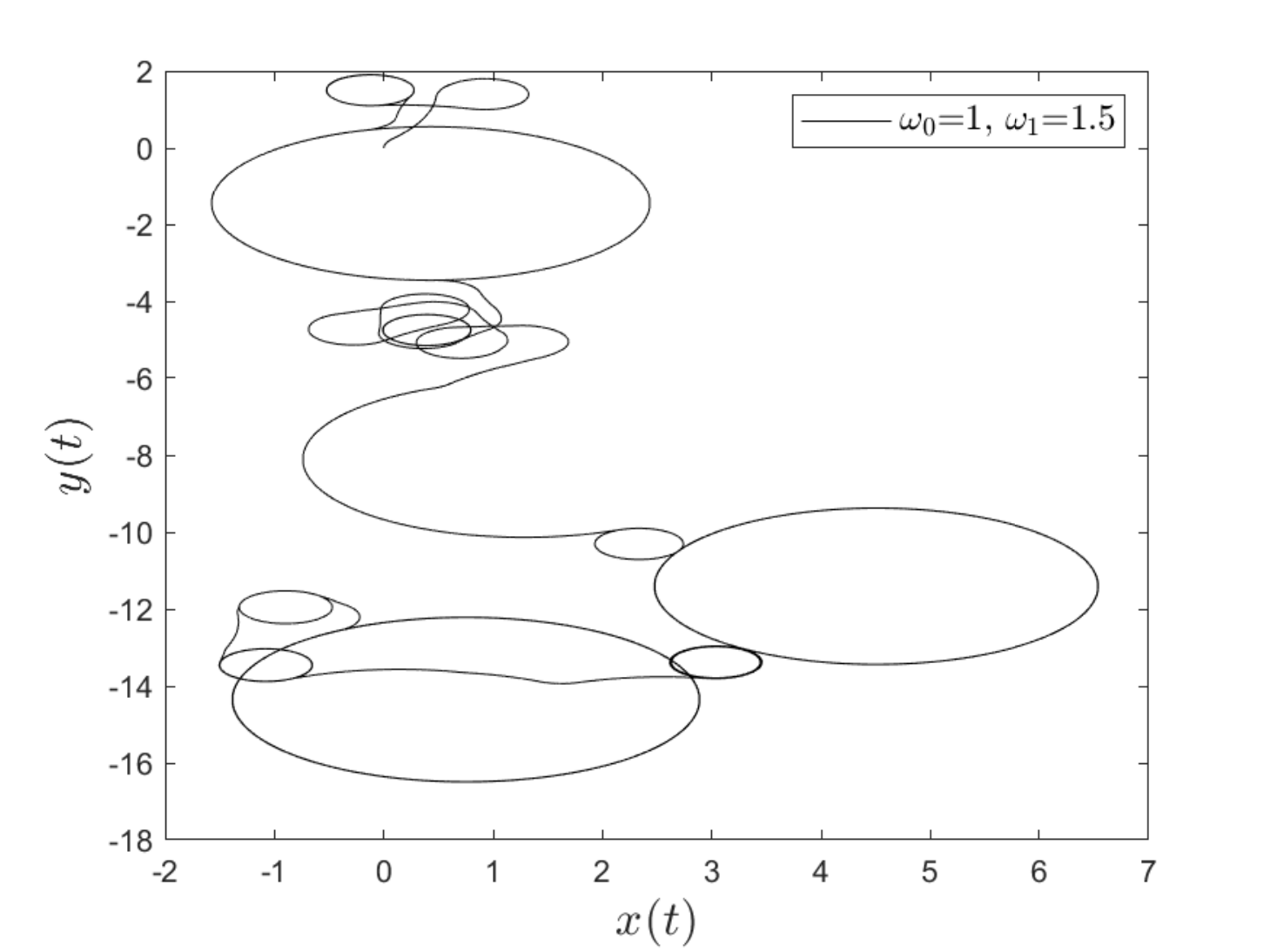}
              \caption{Single realization of the trajectory of a charge in the non-Poisson case. The values of the parameters are $\alpha=0.5$, $T=0.001$,
              $\omega_0=1$, $\omega_1=1.5$ and $t=500$\label{fignopoi2}}
          \end{figure}
  \end{minipage}
   \end{minipage}

\begin{figure}[H]
\includegraphics[width=10 cm]{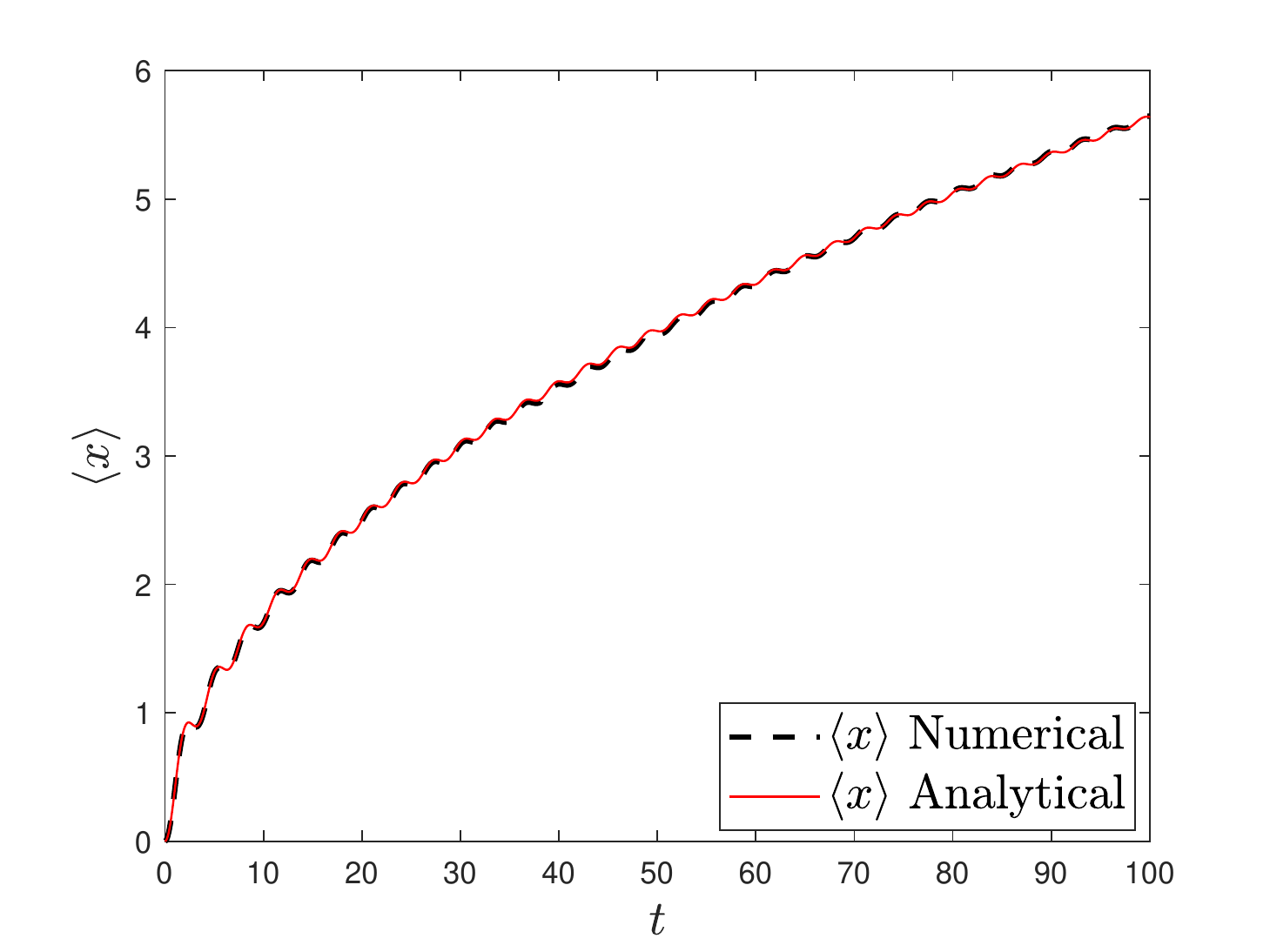}
\caption{Plot of the analytical (redline)  and the numerical solution of the average position $\langle x\rangle$ with $\alpha=0.5$, $T=0.001$, $\omega_0=\omega_1=1$, and $x_0=0$. Number of realizations $200$k. \label{fig3}}
\end{figure}

\begin{figure}[H]
\includegraphics[width=10 cm]{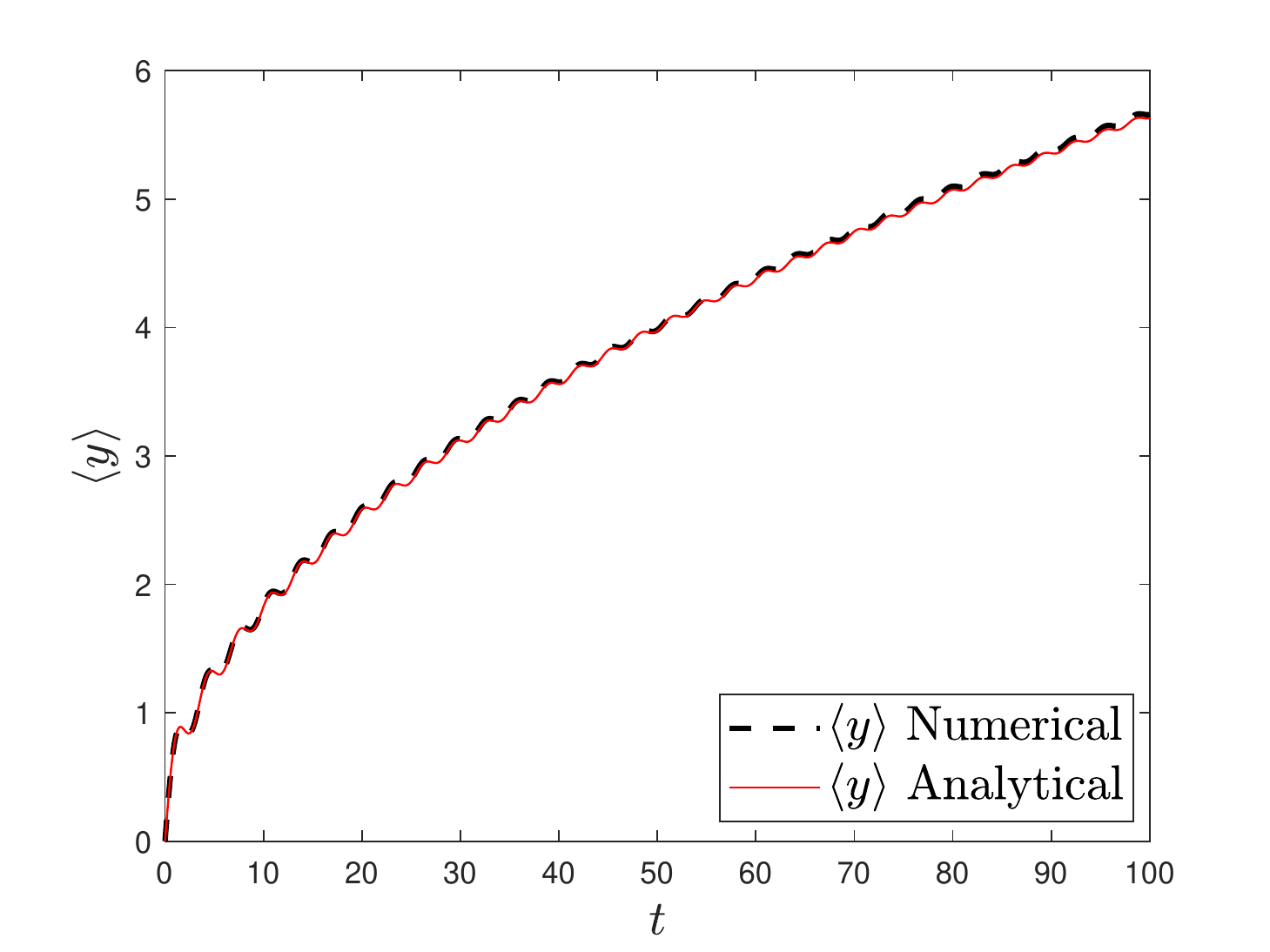}
\caption{Plot of the analytical (redline)  and the numerical solution of the average position $\langle y\rangle$ with $\alpha=0.5$, $T=0.001$, $\omega_0=\omega_1=1$, and $y_0=0$. Number of realizations $200$k. \label{fig4}}
\end{figure}

\begin{figure}[H]
\includegraphics[width=10 cm]{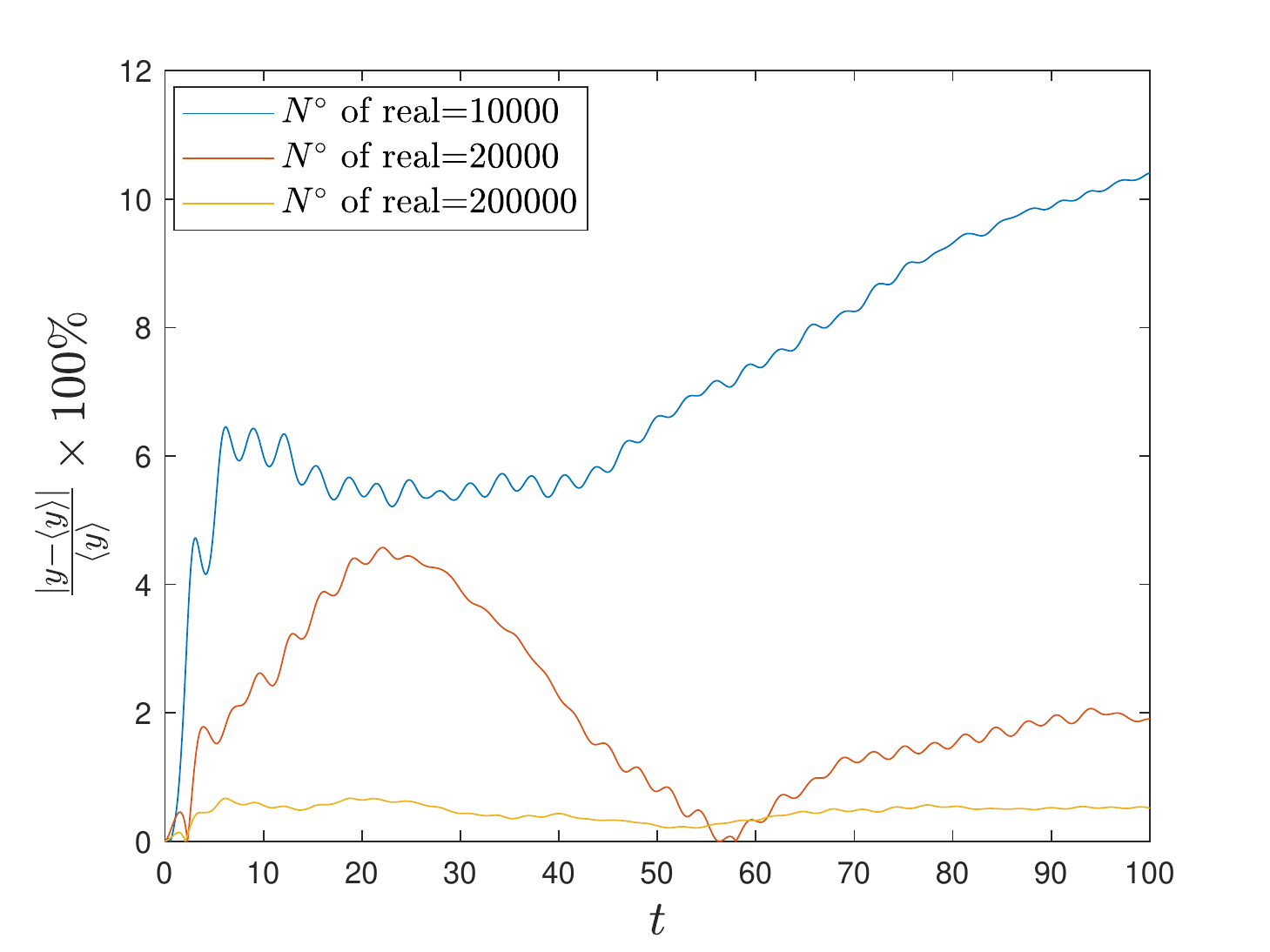}
\caption{Plot of the percent error. On the $x$ axis the time while on the $y$ axis the percentage error between numerical and analytical solution. The different curves are obtained using different number of realizations. \label{figerror}}
\end{figure}

\section{Diffusion}\label{sec_diffusion}
\begin{figure}[H]
\includegraphics[width=10 cm]{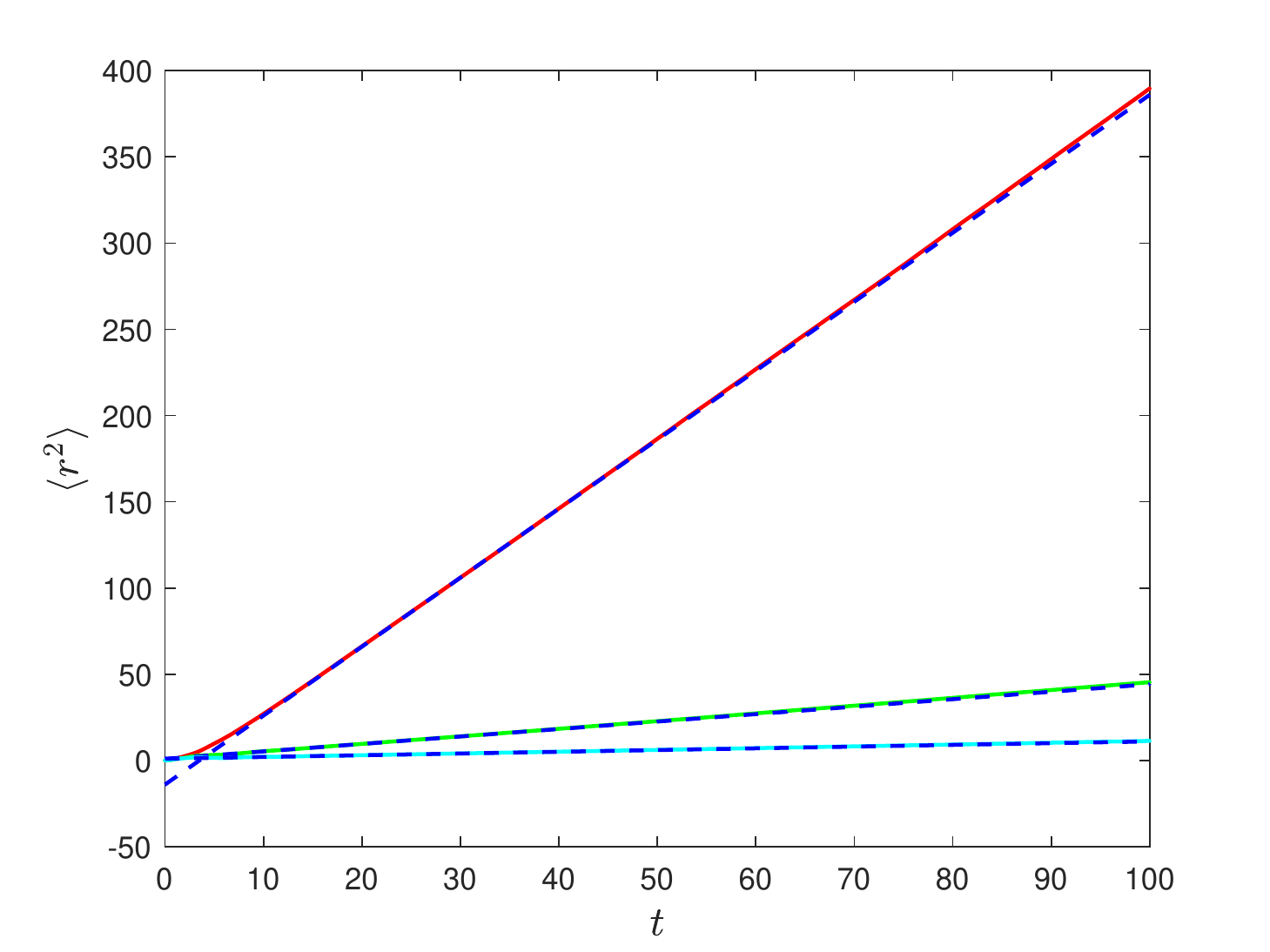}
\caption{Plots    of $\langle r^2\rangle$  in the Poisson case as derived from Eq. \ref{dif4}  compared with numerical simulation. Red continuous line  and blue dashed line refer to the analytical and numerical derivation respectively for parameter values $\gamma=0.5$, $\omega_0=\omega_1=1$ .
Green continuous line and blue dashed line are the  analytical and numerical derivation  for parameters  $\gamma=0.5$, $\omega_0=1,\,\,\omega_1=2$.
Cyan  continuous line and blue dashed line are analytical and numerical derivation for  $\gamma=0.5$, $\omega_0=2,\,\,\omega_1=1$. Number of realizations $200$k.\label{fig1_diff}}
\end{figure}

\begin{figure}[H]
\includegraphics[width=10 cm]{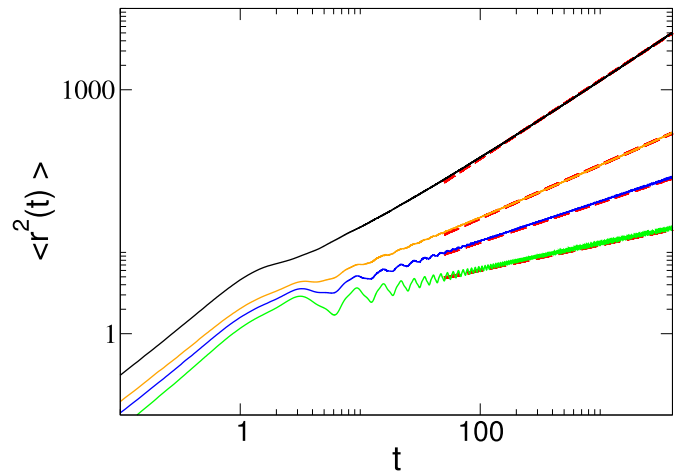}
\caption{Plots of mean-squared displacement $\langle r^2(t)\rangle$  as a function of time. Black, orange, blue and green  line correspond respectively  to $\alpha=1.5, 0.7,0.5, 0.3$. We have chosen $\omega_0=2$ and $\omega_1=1$ and for the waiting time distribution $\psi(t) \propto (t/T)^{-\alpha-1}$ with $T=0.001$.
% , $\alpha=0.7$. $\alpha=0.3$  $\alpha=0.5$.
 Superimposed red dashed  lines
represent guide for the eye for the asymptotic behavior as derived in the text, i.e. $\langle r^2(t)\rangle \propto t $ for $1<\alpha<2$ ( black line case) and $\langle r^2(t)\rangle \propto t^\alpha$ for $0<\alpha<1$ (all other plots).
 \label{r2np}}
\end{figure}

In this section we will evaluate $\langle r^2\rangle=\langle x^2+y^2\rangle$ in the possonian and non-Poisson case. Using Eqs.~(\ref{f3}), (\ref{f4}) we have

\begin{eqnarray}   \label{dif1}
&&x^2(t)=\int_0^t\int_0^t v_x(t_1)v_x(t_2)dt_1dt_2=v^2_0\int_0^t\int_0^t \sin \phi_{t_1}\sin \phi_{t_2}dt_1dt_2
\\ \label{dif2}
&&y^2(t)=\int_0^t\int_0^t v_y(t_1)v_y(t_2)dt_1dt_2=v^2_0\int_0^t\int_0^t \cos \phi_{t_1}\cos \phi_{t_2}dt_1dt_2
\end{eqnarray}
where we set $$\phi_{t}=\int_0^{t} \omega(u) du=\omega_0 t+\omega_1\int_0^t \xi(u)du.$$ Consequently

\begin{equation}\label{dif3}
x^2+y^2=r^2=v^2_0\int_0^t\int_0^t \cos( \phi_{t_1}-\phi_{t_2})dt_1dt_2.
\end{equation}
Our goal is to evaluate the quantity $\langle r^2\rangle$. For sake of compactness let us define the complex quantity

\begin{equation}\label{dif_poiss_def}
r_2(t)=v^2_0\int_0^t\int_0^t \exp\left[i( \phi_{t_1}-\phi_{t_2})\right]dt_1dt_2,
\end{equation}
and take its time derivative

\begin{eqnarray}\label{aqui_nonpoiss0}
\frac{\partial}{\partial t}r_2(t)=2 v^2_0 \int_0^t \exp\left[i\omega_0( t-t_1)\right] \Big\langle\exp\left[ i\omega_1\int_{t_1}^t \xi(u)du\right]\Big\rangle dt_1. 
\end{eqnarray}
 {For a Poisson process,  with exponential waiting times distribution and correlation, the distribution of the first observed jump/event is the same as that of any other following event \cite{aq2}, this means that when averaging over the fluctuations, shifting the time origin, will not affect the result.  For a generic non-Poissonian process, but with a finite time scale, this remains a good approximation in the long-time limit,
%When the process has a finite time scale we may approximate 
so that  we may re-write Eq.~(\ref{aqui_nonpoiss0}) as}

\begin{eqnarray}\label{aqui_nonpoiss1}
\frac{\partial}{\partial t}r_2(t)\approx 2 v^2_0 \int_0^t \exp\left[i\omega_0( t-t_1)\right] \Big\langle\exp\left[ i\omega_1\int_{0}^{t-t_1} \xi(u)du\right]\Big\rangle dt_1.
\end{eqnarray}
Formally, for $t\to\infty$, Eq.~(\ref{aqui_nonpoiss1}) is the Laplace transform of the averaged function where $s\to i\omega_0$. Using the result of Eq.~(\ref{general}), we may write

\begin{eqnarray}\nonumber
&&\frac{\partial}{\partial t}r_2(t)\approx  v^2_0 \left[\frac{\left(1+\hat{\psi} \left(i\omega_0-i \omega _1\right)\right) \hat{\Psi} \left(i\omega_0+i \omega _1\right)}{1-\hat{\psi} \left(i\omega_0-i \omega _1\right) \hat{\psi}\left(i\omega_0+i \omega _1\right)}+\right.
\\\label{aqui_nonpoiss2}
&&\left.\frac{\left(1+\hat{\psi} \left(i\omega_0+i \omega _1\right)\right) \hat{\Psi} \left(i\omega_0-i \omega _1\right)}{1-\hat{\psi}\left(i\omega_0-i \omega _1\right)\hat{\psi} \left(i\omega_0+i \omega _1\right)}\right],
\end{eqnarray}
namely a constant. We deduce that for the Poisson and the non-Poisson case, but with $1<\alpha<2$, we have ordinary diffusion, i.e.

\begin{eqnarray}\label{r2finale}
\langle r^2\rangle=\mathrm{Re}[r_2(t)]\propto t
\end{eqnarray}
 {Note that in the Poisson case the approximation~(\ref{aqui_nonpoiss1}) is actually an exact expression and~$\Big\langle\exp\left[ i\omega_1\int_0^{t-t_1} \xi(u)du\right]\Big\rangle$ is given by Eq. (\ref{average_eq}). The integration of~(\ref{aqui_nonpoiss1}) and the subsequent integration to obtain $\langle r^2\rangle$ does not present difficulties being the integral functions exponential functions. Neglecting the transient due to the exponentials with negative real part, the asymptotic expression for $\langle r^2\rangle$ reads as}

\begin{equation}\label{dif4}
\langle r^2\rangle\approx \frac{2 \gamma    v_0^2 \omega _1^2}{\gamma ^2 \omega _0^2+\left(\omega _0^2-\omega _1^2\right)^2}t,\,\,t\to\infty.
\end{equation}
The diffusion is faster at resonance, when $\omega_0=\omega_1$. The numerical check for the quantity $\langle r^2\rangle$, obtained  integrating the uniform circular motion between switches of the magnetic field value, is shown in Figures~\ref{fig1_diff} (Poisson case) and \ref{r2np} ($1<\alpha<2$). The agreement with the analytical result, Eq. (\ref{r2finale}), is remarkable.

When~$0<\alpha<1$ there is not a finite time scale, and we can not use the approximation given by Eq~(\ref{aqui_nonpoiss1}).  {It is important to notice that  while in the  poissonian case the  distribution of the first observed event/jump coincides with the distribution $\psi(t)$ of any other event, in the non-poissonian case it will be different from $\psi(t)$ and it will be function also of the time $t_a<0$, the time at which the system is prepared, in other words, the distribution  of the first observed event will have a two-times dependence $\psi(t,t_a)$}. The distribution of the first jump has to be considered, and we are forced to use the full formula of Ref.~\cite{aq2}, i.e.

\begin{eqnarray}\nonumber
&&{\cal L}\left[ \int_0^t \exp\left[i\omega_0( t-t_1)\right] \Big\langle\exp\left[ i\omega_1\int_{t_1}^t \xi(u)du\right]\Big\rangle dt_1 \right]=
 \\\nonumber
&&\frac{1}{2}  \frac{\left[\hat{f}_{-}(s)+\hat{f}_{+}(s)\hat{\psi} \left(s- i\delta_-\right)\right]\hat{\Psi} \left(s-i\delta_+\right)}{1-\hat{\psi} \left(s- i \delta_+\right)\hat{\psi} \left(s- i \delta_-\right)} +\frac{1}{2}\hat{F}_+ \left(s\right)+
 \\\label{aqui_nonpoiss}
 &&\frac{1}{2} \frac{\left[\hat{f}_{+}(s)+\hat{f}_{-}(s)\hat{\psi} \left(s- i\delta_+\right)\right]\hat{\Psi} \left(s-i\delta_-\right)}{1-\hat{\psi} \left(s- i \delta_+\right)\hat{\psi} \left(s- i \delta_-\right)} 
 +\frac{1}{2}\hat{F}_- \left(s\right)
\end{eqnarray}
where $\delta_{\pm}=\omega_0\pm\omega_1$, and

\begin{equation}\label{first_jump}
\hat{f}_{\pm}(s)={\cal L}\left[\int_{0}^t f(t-t_1,t_1)\exp[i\delta_{\pm}(t-t_1)]dt_1\right]=
 \frac{\hat{\psi} \left(s\right)-\hat{\psi} \left(s-i\delta_{\pm}\right)}{-i\delta_{\pm}\left[1-\hat{\psi} \left(s- i \delta_+\right)\right]}
\end{equation}
being $f(\tau,t_1)$ the conditional probability density that, fixed a time $t_1$, the first next switching event of the variable $\xi(t)$ occurs at time $t_1+\tau$. It is important to notice that  differently from the Poisson case, this distribution is different from the distribution $\psi(t)$ of any other event. It coincides with $\psi$ only for the Poisson case. Analogously

\begin{equation}\label{first_jumpF}
\hat{F}_{\pm}(s)={\cal L}\left[\int_{0}^t F(t-t_1,t_1)\exp[i\delta_{\pm}(t-t_1)]dt_1\right]= \frac{1/s-\hat{f}_{\pm}(s)}{s-i\delta_{\pm}}
\end{equation}
where $$F(\tau,t_1)=1-\int_{0}^\tau f(\tau',t_1)d\tau'$$ is the conditional probability that, fixed $t_1$, no switch occurs between $t=t_1$ and $t=t_1+\tau$. Using distribution (\ref{nd11}) or in alternative the Mittag-Leffler distribution, we may write for the inverse Laplace transform of $\hat{f}_{\pm}(s)$ the following

\begin{eqnarray}\nonumber
\label{primosalto}
&&f_{\pm}(t)={\cal L}^{-1}\left[\hat{f}_{\pm}(s)\right]=-  
 \frac{1}{\delta_{\pm}}\frac{d}{dt}E_\alpha\left[-\left(\frac{t}{T}\right)^{\alpha} \right]-\frac{1}{\delta_{\pm}T}\frac{1}{\Gamma(\alpha)}  \left(\frac{t}{T}\right)^{\alpha-1}-
 \\\label{aqui_nonpoiss_distr}
&&  \frac{1}{\delta_{\pm}T\Gamma(\alpha)}
\int_0^t \exp\left[i\delta_{\pm}( t-t_1)\right]   \left(\frac{t-t_1}{T}\right)^{\alpha-1}       
\frac{d}{dt_1}E_\alpha\left[-\left(\frac{t_1}{T}\right)^{\alpha}\right]  dt_1
\end{eqnarray}
 {From Eq. (\ref{primosalto}), one can also understand why the contribution of the first jump/event becomes crucial for a non-Poissonian process for $0<\alpha<1$.
In this regime, differently from the Poisson regime, not only is such distribution different from that of any following event \cite{aq2}, but it also becomes dominant asymptotically.} %compared to the case where  other event,  since it  decays as $~t^{\alpha-1}$ instead of $t^{-\alpha -1}.}
Analyzing Eq.~(\ref{first_jumpF}) from an asymptotic point of view, namely $sT\to 0$ and $\delta T\ll 1$ we deduce that contribution to the diffusion is given by $F_{\pm}$.

\begin{eqnarray}\label{Fint}
F_{\pm}(t)=\frac{i}{\delta_{\pm}}-\frac{i \exp[i\delta_{\pm} t] }{\delta_{\pm}}+\int_0^t \exp[i\delta_{\pm} (t-t_1)]f(t_1)dt_1
\end{eqnarray}
Being $f_{\pm}(t)$ a function decaying with time we may approximate $F_{\pm}(t)$ for $t\to\infty$ (see Ref.~\cite{bol3} for more details)

\begin{eqnarray}\label{Fint2}
F_{\pm}(t)\approx \frac{i}{\delta_{\pm}}-\frac{i \exp[i\delta_{\pm} t] }{\delta_{\pm}}+\exp[i\delta_{\pm} t] \hat{f}_{\pm}(i\delta_{\pm}).
\end{eqnarray}
The contribution of $F_{\pm}(t)$ to $\langle r^2(t)\rangle $ generates constant and oscillating terms while the dominant term is a power law. Indeed, we have

\begin{eqnarray}\label{Final}
\langle r^2(t)\rangle \sim \textrm 2 v_0^2{Re}[A]\frac{1}{\Gamma[\alpha] \Gamma[\alpha-1]}\left(\frac{t}{T}\right)^{\alpha},\,\,\mathrm{for}\,\,t\to\infty.
\end{eqnarray}
The analytical result is obtained expanding for small $s$ the Laplace transform of the expression for the derivative of $\langle r^2(t)\rangle$, neglecting the terms $\hat{F}_{\pm}(s)$ for the reason stated above. The following expression gives the coefficient $A$

%\begin{eqnarray}\nonumber
%&&A= \frac{1}{1-\hat{\psi}_+\hat{\psi}_-}\left[ \frac{\hat{\Psi}_+(1-\hat{\psi}_+)\hat{\psi}_-}{\delta_+}+ \frac{\hat{\Psi}_+(1-\hat{\psi}_-)}{\delta_-}+\frac{\hat{\Psi}_-(1-\hat{\psi}_-)\hat{\psi}_+}{\delta_-}+\right. 
%\\\nonumber
% &&\left.+ \frac{\hat{\Psi}_-(1-\hat{\psi}_+)}{\delta_+}\right]+\frac{1-\hat{\psi}_+}{\delta_+^2}+\frac{1-\hat{\psi}_+}{\delta_-^2} 
%\end{eqnarray}
\begin{equation}
A=\frac{1}{1-\hat{\psi}_+\hat{\psi}_-}\left[ \frac{\hat{\Psi}_+(1-\hat{\psi}_+)\hat{\psi}_-}{i\delta_+}+ \frac{\hat{\Psi}_+(1-\hat{\psi}_-)}{i\delta_-}\right] -\frac{1-\hat{\psi}_+}{\delta_+^2} +  \left(+\leftrightarrow - \right)
\end{equation}
with $\hat{\Psi}_{\pm}$ and $\hat{\psi}_{\pm}$ the Laplace transform of the respective functions $\Psi(t)$ and $\psi(t)$ evaluated in $s=-i \delta_{\pm}$ and the last term within round brackets is obtained from previous terms by exchanging $+$ and $-$ subscripts. The numerical check  is shown in Figure~ \ref{r2np}

%%%%%%%%%%%%%%%%%%%%%%%%%%%%%%%%%%%%%%%%%%
\section{Results}\label{sec_conclusions}
To summarize, we have introduced a complete framework to describe the motion of a charged particle in a fluctuating magnetic field.
We considered both ergodic and non-ergodic fluctuations of the magnetic field. We find that in the case of ergodic fluctuations
the diffusion is asymptotically normal, while for non-ergodic fluctuations, we find anomalous diffusion properties.
The diffusion is characterized by the mean-squared displacement, which we derived analytically and confirmed numerically for both regimes as illustrated in Figures \ref{fig1_diff} and  \ref{r2np}.
In the case of non-Poisson fluctuations, we provide analytical formulas in Laplace transform from which we extract the asymptotic time behavior for the mean-squared displacement.

%\subsection{Subsection}
%subsubsection{Subsubsection}

%%%%%%%%%%%%%%%%%%%%%%%%%%%%%%%%%%%%%%%%%%
\section{Discussion and conclusions}\label{sec_conclusions2}
The problem of the motion of a charged particle in fluctuating magnetic field has not been investigated when fluctuations of the field have non-ordinary statistical properties despite its physical interest and possible applications e.g. to plasma models.
This paper starts filling this gap by considering dichotomic fluctuations with both Poisson and non-Poisson properties.  {For the Poisson case, we find the driven equation for the probability density. On the contrary, the equation for probability density function in the non-ordinary statistics case is still an open problem.
We developed a theoretical framework for the first and second moment of the particle position and afforded both an analytical and numerical description. We neglected in this framework the effect of induced electric field due to the variations of the magnetic field, which we leave out for further investigation in an upcoming publication.} 
Interestingly we find that diffusion is normal either when fluctuations are Poisson or non-Poisson but with a finite mean time, differently from the standard case of  continuous-time random  walk (CTRW) which shows anomalous diffusion behavior for power-law distribution with finite mean-time (either in the velocity or jump model)\cite{zumofen}.
When the fluctuations time scale diverges, i.e., for non-ergodic fluctuations, an anomalous diffusion  regime emerges, again differently from standard CTRW
where a  ballistic regime applies  for power-law distributions with diverging mean-time.  {This difference is related to the fact that the length of the CTRW jumps is proportional to the elapsed time while, In the case studied in the paper, the particle is forced to move along a circular trajectory. Thus the jumps can not exceed the radius of the circumference, causing the relevant differences between the two diffusion processes.}

%%%%%%%%%%%%%%%%%%%%%%%%%%%%%%%%%%%%%%%%%%
\vspace{6pt} 

%%%%%%%%%%%%%%%%%%%%%%%%%%%%%%%%%%%%%%%%%%
%% optional
%\supplementary{The following are available online at \linksupplementary{s1}, Figure S1: title, Table S1: title, Video S1: title.}


%%%%%%%%%%%%%%%%%%%%%%%%%%%%%%%%%%%%%%%%%%
\authorcontributions{G.A. performed research, did computational simulations. K.J.C. performed computational simulations, did calculations. M.B. did calculations, performed derivations. The manuscript was prepared by all authors. All authors have read and agreed to the published version of the manuscript.
}

\funding{This research received no external funding}

\institutionalreview{Not applicable.}

\informedconsent{Not applicable.}

\dataavailability{Data sharing not applicable.} 

%\acknowledgments{In this section you can acknowledge any support given which is not covered by the author contribution or funding sections. This may include administrative and technical support, or donations in kind (e.g., materials used for experiments).}

\conflictsofinterest{The authors declare no conflict of interest.} 

%% Optional
%\sampleavailability{Samples of the compounds ... are available from the authors.}

%%%%%%%%%%%%%%%%%%%%%%%%%%%%%%%%%%%%%%%%%%
%% Only for journal Encyclopedia
%\entrylink{The Link to this entry published on the encyclopedia platform.}

%%%%%%%%%%%%%%%%%%%%%%%%%%%%%%%%%%%%%%%%%%
%% Optional

%%%%%%%%%%%%%%%%%%%%%%%%%%%%%%%%%%%%%%%%%%
%% Optional
\appendixtitles{no} % Leave argument "no" if all appendix headings stay EMPTY (then no dot is printed after "Appendix A"). If the appendix sections contain a heading then change the 

%%%%%%%%%%%%%%%%%%%%%%%%%%%%%%%%%%%%%%%%%%
\end{paracol}
\reftitle{References}

\end{document}